\documentclass[prl, aps, amssymb, amsmath, nofootinbib, twocolumn]{revtex4}
	
	\usepackage{graphicx}
	\usepackage{color}
	\usepackage{verbatim}
	\usepackage{subfigure}
	\include{subcaption}
	\usepackage{amssymb}

\begin{document}

	\title{Analysis of slow transitions between nonequilibrium steady states}

	\author{Dibyendu Mandal$^1$ and Christopher Jarzynski$^2$	}
	\affiliation{ 
	$^1$Department of Physics, University of California, Berkeley, CA 94720, U.S.A.\\
	$^2$Department of Chemistry and Biochemistry and Institute for Physical Science and Technology, University of Maryland, College Park, MD 20742, U.S.A.
	}

	\begin{abstract}	
	
	Transitions between nonequilibrium steady states obey a generalized Clausius inequality, which becomes an equality in the quasistatic limit.
	For slow but finite transitions, we show that the behavior of the system is described by a response matrix whose elements are given by a far-from-equilibrium Green-Kubo formula, involving the decay of correlations evaluated in the nonequilibrium steady state.
	This result leads to a fluctuation-dissipation relation between the mean and variance of the nonadiabatic entropy production, $\Delta s_{\rm na}$.
	Furthermore, our results extend -- to nonequilibrium steady states -- the thermodynamic metric structure introduced by Sivak and Crooks for analyzing minimal-dissipation protocols for transitions between equilibrium states.
		
	\end{abstract}

	\maketitle

	Classical thermodynamics is a macroscopic theory built around the concept of the equilibrium state~\cite{Callen}, whereas statistical mechanics is a microscopic theory that represents equilibrium states in terms of known statistical ensembles~\cite{Peliti}.
	Both frameworks accurately describe the equilibrium properties of matter and the constraints that must be satisfied when systems undergo transitions between equilibrium states.
	No comparably general theories exist for systems away from equilibrium, not even for steady states.
	We have no {\it a priori} microscopic representations of nonequilibrium steady states; neither do we have a macroscopic, empirical thermodynamic theory to compare to. 

	A phenomenological thermodynamic theory of nonequilibrium steady states was suggested by Oono and Paniconi~\cite{Oono1998}, who introduced the crucial concept of {\it housekeeping heat}, which is the heat transfer necessary to maintain a system in a given steady state.
	The {\it excess heat} $Q_\text{ex}$ is then the renormalization of the total heat $Q$ by the housekeeping heat $Q_\text{hk}$:
	\begin{equation}
	\label{eq:Excess}
	Q_\text{ex}  =  Q - Q_\text{hk}. 
	\end{equation}
	In the context of Markovian dynamics, these notions were modified somewhat and given quantitative expressions by Hatano and Sasa~\cite{Hatano2001}, Speck and Seifert~\cite{Speck2005} and Ge and Qian~\cite{Ge2009,Ge2010}, leading to derivations of a generalized Clausius inequality for transitions between steady states~\cite{Hatano2001, Ge2009}:
	\begin{equation}
	\label{eq:mClausius}
	\Delta S + \int \mathrm{d}t \, \beta(t) \langle \dot{Q}_\text{ex}\rangle  \geq 0. 
	\end{equation}
	Here $\Delta S$ denotes the net change in the entropy of the system, and $\langle \dot{Q}_\text{ex}\rangle$ is the ensemble-averaged rate of excess heat transfer to a reservoir at inverse temperature $\beta(t)$.
	(See also Bertini {\it et al}~\cite{Bertini2012,Bertini2013} for a related inequality for renormalized {\it work}.)
	Under equilibrium dynamics Eq.~\ref{eq:mClausius} reduces to the usual Clausius inequality, as $Q_\text{hk}=0$ and hence $Q_\text{ex}=Q$.
			
	Eq.~\ref{eq:mClausius} suggests that there may exist a steady-state thermodynamic framework that closely parallels equilibrium thermodynamics, with steady states and excess heat in the roles traditionally assigned to equilibrium states and heat, respectively.
   	In this Letter we further develop these parallels, by analyzing {\it finitely slow} transitions from one nonequilibrium steady state to another.
	At leading order in a perturbative expansion in the driving speed, Eq.~\ref{eq:mClausius} becomes an equality and hysteresis vanishes, suggesting that quasistatic steady-state processes are natural counterparts of reversible thermodynamic processes~\cite{Oono1998}.
	At the next order, a response matrix $\zeta$ governs the renormalized entropy production (Eq.~\ref{eq:Slow2}).
	We show that the elements $\zeta_{\mu\nu}$ are given by time-integrated correlation functions evaluated in the steady state (Eq.~\ref{eq:FDT}), in exact analogy with equilibrium Green-Kubo relations.
	These results lead to a fluctuation-dissipation theorem, which states that the average renormalized entropy production is equal to half its variance (Eq.~\ref{eq:Gaussian}).
	Finally, our analysis generalizes recent progress related to thermodynamic length and its applications to the determination of optimal driving protocols~\cite{Salamon1983, Crooks2007, Sivak2012, Zulkowski2012}.
	
	Our results reveal that the stochastic theory of slow transitions between equilibrium states~\cite{Speck2004,Sivak2012,Hoppenau2013} extends directly to slow transitions between nonequilibrium steady states, under the definition of housekeeping heat proposed in Refs.~\cite{Hatano2001,Speck2005,Ge2009,Ge2010}.
	We note that alternative definitions have been suggested, and corresponding generalized Clausius inequalities have been derived, by Komatsu {\it et al}~\cite{Komatsu2008,Komatsu2010} and Maes and Neto\v{c}n\' y~\cite{Maes2014}.
	It remains an open question whether the results we derive have counterparts within the frameworks of Refs.~\cite{Komatsu2008,Komatsu2010,Maes2014}.
		
 	We develop our theory within the context of isothermal Markovian dynamics on a network, in which the system's evolution consists of random transitions among a set of discrete states.
	This mesoscopic level of description is well-suited for small stochastic systems such as molecular motors~\cite{Kolomeisky2007}.
	We expect that our results extend as well to diffusive processes, described by Langevin~\cite{vanKampen2007} and Fokker-Planck equations~\cite{Risken1984}.
	In the following three paragraphs we specify notation and define quantities that will play important roles in our subsequent analysis.

	Consider a system with $N$ discrete states $i \in \{1, 2, \ldots N\}$, in contact with a reservoir at temperature $\beta^{-1}$. 
	Induced by thermal fluctuations from the reservoir, the system makes random, Poissonian transitions among its states, with $R_{ij} \geq 0$ denoting the transition rate from $j$ to $i$.
	The probability distribution ${\bf p} = (p_1, p_2, \ldots p_N)^T$ (the superscript $T$ denotes transposition) satisfies
	\begin{equation}
	\label{eq:Master}
	\frac{d}{dt} {\bf p} = {\cal R} \, {\bf p}, 
	\end{equation}
where the rate matrix ${\cal R}$ is formed by the rates $R_{ij}$. 
	The diagonal elements satisfy $R_{ii} = - \sum_{j \neq i} R_{ji}$ to preserve normalization: $\sum_i p_i(t) = 1$.
	The current
	\begin{equation}
	\label{eq:Current}
	J_{ij} = R_{ij} p_j - R_{ji} p_i
	\end{equation} 
	is the instantaneous flow of probability from $j$ to $i$.
	The quantity $Q_{ij} =  \beta^{-1}\ln{\left(R_{ij}/R_{ji}\right)}$ represents the heat transferred from the system to its thermal surroundings, during a transition from $j$ to $i$~\cite{Seifert2012}.
	Thus the average instantaneous rate of heat transfer to the reservoir is
	\begin{equation}
	\label{eq:Srdot}
	\langle \dot{Q} \rangle =  \beta^{-1} \sum_{i < j} J_{ij} \ln{\left(\frac{R_{ij}}{R_{ji}}\right)}.
	\end{equation}
	
	We further assume that (i) $R_{ij} \neq 0$ if and only if $R_{ji}\neq 0$, and (ii) any state $i$ can be reached from any other state $j$ either directly or via intermediate states. 
	Under these assumptions, any initial distribution ${\bf p}(0)$ relaxes to a unique steady state distribution ${\boldsymbol \pi}$, with ${\cal R} {\boldsymbol \pi} = 0$, characterized by steady currents $J_{ij}^\text{ss} = R_{ij} \pi_j - R_{ji} \pi_i$~\cite{vanKampen2007}.
	If $J_{ij}^\text{ss}=0$ for all $i\ne j$, then the dynamics generated by ${\cal R}$ satisfy detailed balance, and the distribution $\boldsymbol\pi$ represents an equilibrium state.
	If some $J_{ij}^\text{ss} \ne 0$ -- as we generically assume throughout this Letter -- then detailed balance is broken and $\boldsymbol\pi$ specifies a nonequilibrium steady state.
 
	We now define the quantities
	\begin{equation}
	\label{eq:SSForce}
	F^\text{ss}_{ij} = \ln{\left(\frac{R_{ij} \pi_j}{R_{ji} \pi_i}\right)},
	\end{equation} 
which we interpret as effective thermodynamic forces that drive the probability currents in nonequilibrium steady states (note that $F^\text{ss}_{ij} = 0$ if and only if $J^\text{ss}_{ij} = 0$).
	These forces are uniquely determined by the rate matrix ${\cal R}$, and are nonlocal, in the sense that each $F_{ij}^\text{ss}$ generally depends on all the elements of ${\cal R}$, via the stationary distribution ${\boldsymbol \pi}$.
	The instantaneous rate of housekeeping heat transfer to the medium is now defined as~\cite{Ge2009,Ge2010}
	\begin{equation}
	\label{eq:Housekeeping}
	\langle \dot{Q}_\text{hk} \rangle =  \beta^{-1}  \sum_{i < j} J_{ij} F^\text{ss}_{ij} \, ,
	\end{equation}
which has a natural interpretation as the power associated with the forces $F^\text{ss}_{ij}$. 
	It can be shown that $\langle \dot{Q}_\text{hk} \rangle \ge 0$ for any distribution ${\bf p}$, and $\langle \dot{Q} \rangle = \langle \dot{Q}_\text{hk} \rangle$ in the steady state ${\bf p} = {\boldsymbol \pi}$~\cite{Ge2009, Ge2010}. 
Using Eq.~\ref{eq:Excess}, we obtain the excess heat transfer rate,
	\begin{equation}
	\label{eq:ExcessRate}
	\langle \dot{Q}_\text{ex} \rangle = \langle \dot{Q} \rangle - \langle \dot{Q}_\text{hk} \rangle = \beta^{-1} \sum_{i < j} J_{ij} \ln{\left(\frac{\pi_i}{\pi_j}\right)}. 
	\end{equation}

	Consider now a set of parameters ${\boldsymbol \lambda} = \{\lambda_1, \lambda_2, \ldots \lambda_K\}$ that determine the transition rates $R_{ij}({\boldsymbol \lambda})$ and corresponding steady states ${\boldsymbol \pi}^{\boldsymbol\lambda}$.
	The generalized Clausius inequality, Eq.~\ref{eq:mClausius}, applies to processes in which the system is driven from state ${\boldsymbol \pi}^{\bf A}$ to state ${\boldsymbol \pi}^{\bf B}$, by varying these parameters from ${\bf A}$ to ${\bf B}$ over a time interval $\Delta t$.
	We assume that the protocol ${\boldsymbol \lambda}(t)$ is smooth, so that $d{\boldsymbol \lambda}/dt$ is well-defined. 
	The term $\Delta S$ appearing in Eq.~\ref{eq:mClausius} is the change in Shannon entropy, $S({\boldsymbol \pi_{\bf B}}) - S({\boldsymbol \pi_{ \bf A}})$, with $S = - \sum_i \pi_i \ln{\pi_i}$. 
	
	We will analyze Eq.~\ref{eq:mClausius} for processes in which the parameters are varied {\it slowly}, hence the system remains near the nonequilibrium steady state. 
	For convenience, we introduce a small parameter $\epsilon\propto \vert d{\boldsymbol \lambda}/dt \vert$, so that $\Delta t\propto\epsilon^{-1}$ for fixed ${\bf A}$ and ${\bf B}$.
	We will find that in the quasistatic limit ($\epsilon\rightarrow 0$) as well as in the leading correction to this limit, the response of the system is remarkably analogous to that of a slowly driven equilibrium system.
		
	Let us write ${\bf p}(t)$ as the sum of the instantaneous steady state distribution ${\boldsymbol \pi}(t) = {\boldsymbol \pi}^{{\boldsymbol\lambda}(t)}$ and a small correction, or ``lag'', $\Delta {\bf p}(t)$:
	\begin{equation}
	\label{eq:Decomposition}
	{\bf p}(t) = {\boldsymbol \pi}(t) + \Delta {\bf p}(t).
	\end{equation}
	Substituting Eq.~\ref{eq:Decomposition} into Eq.~\ref{eq:Master} and using the condition ${\cal R} {\boldsymbol \pi} = {\bf 0}$, we obtain the linear inhomogeneous equation,
	\begin{equation}
	\label{eq:MasterCorr}
	\frac{d}{dt} \Delta {\bf p} - {\cal R} \Delta {\bf p} = - \frac{d}{dt} {\boldsymbol \pi}.
	\end{equation}
	Let us now define a generalized inverse ${\cal R}^+$ by the relations~\cite{Boullion1971, Horowitz2010}
	\begin{subequations}
	\begin{eqnarray}
	\label{eq:MPInverse}
	\left( {\cal R} {\cal R}^+ \right)_{ij} &=& \left( {\cal R}^+ {\cal R} \right)_{ij} = \delta_{ij} - \pi_i \\
	{\cal R^+}{\boldsymbol\pi} &=& {\bf 0} \quad,\quad {\bf 1}^T {\cal R^+} = {\bf 0}^T ,
	\end{eqnarray}
	\end{subequations}
where $\delta_{ij}$ is the Kronecker delta, ${\bf 0}^T = (0,0,\cdots 0)$, and ${\bf 1}^T = (1,1,\cdots 1)$.
	Applying ${\cal R}^+$ to both sides of Eq.~\ref{eq:MasterCorr}, we get
	\begin{equation}
	\label{eq:Basis}
	\left[ 1 - {\cal R}^+ \frac{d}{dt} \right] \Delta {\bf p} = {\cal R}^+ \frac{d}{dt} {\boldsymbol \pi},
	\end{equation}
	which is solved iteratively to obtain 
	\begin{equation}
	\label{eq:Series}
	\Delta {\bf p} = \sum_{n = 1}^\infty \left( {\cal R}^+ \frac{d}{dt} \right)^n {\boldsymbol \pi} \equiv \sum_{n=1}^\infty {\bf a}_n \epsilon^n .
	\end{equation}
	This expansion of the lag $\Delta{\bf p}$ in powers of the driving rate leads to a corresponding expansion $\langle \dot{Q}_\text{ex} \rangle = \sum_{n=1}^\infty b_n \epsilon^n$ (via Eq.~\ref{eq:mExcess} below).
	In the following, we examine in turn the first two terms of this expansion.

	We start by rewriting Eq.~\ref{eq:ExcessRate} in the form~\cite{Ge2009, Ge2010}
	\begin{equation}
	\label{eq:mExcess}
	\langle \dot{Q}_\text{ex} \rangle = \beta^{-1}\sum_{i, j} R_{ij} p_j \ln{\pi_i}.
	\end{equation}
	From Eq.~\ref{eq:Decomposition} and the $n = 1$ term of Eq.~\ref{eq:Series} we get
	\begin{equation}
	\label{eq:Quasistatic}
	\beta \langle \dot{Q}_\text{ex} \rangle = \sum_{i, j} R_{ij} \pi_j \ln{\pi_i} + \sum_{i, j, k} R_{ij} R^+_{jk} \dot{\pi}_k \ln{\pi_i}. 
	\end{equation} 
	The first term on the right vanishes, since ${\cal R} {\boldsymbol \pi} = {\bf 0}$. 
	Using Eq.~\ref{eq:MPInverse} in the second term, we get a sum of two terms: (1) $\sum_{i} \dot{\pi}_i \ln{\pi_i}$, which is equal to $-dS/dt$, and (2) $-\sum_{i, k} \pi_{i} \dot{\pi}_k \ln{\pi_i}$, which vanishes by conservation of normalization: $\sum_k \dot{\pi}_k = 0$.
	We thus arrive at
	\begin{equation}
	\label{eq:reversible}
	\beta \langle \dot{Q}_\text{ex} \rangle = - \frac{dS}{dt} + {\cal O}(\epsilon^2),
	\end{equation}
	which implies that in the quasistatic limit ($\epsilon\rightarrow 0$, with $\Delta t\propto\epsilon^{-1}$), Eq.~\ref{eq:mClausius} becomes an equality:
	\begin{equation}
	\label{eq:mClausius_rev}
	\Delta S + \int \mathrm{d}t \, \beta(t) \langle \dot{Q}_\text{ex}\rangle \stackrel{qs}{=} 0 
	\end{equation}
	This result is a generalized Clausius {\it equality} for quasistatic transitions between nonequilibrium steady states.
	An equivalent result was obtained for overdamped Langevin processes in Ref.~\cite{Hatano2001, Maes2014}.
	Eq.~\ref{eq:mClausius_rev} implies that the integral $\int \mathrm{d}t \, \beta(t) \langle \dot{Q}_\text{ex}\rangle$ is independent of the quasistatic path taken from ${\bf A}$ to ${\bf B}$ in $\boldsymbol\lambda$-space, and therefore vanishes when the path is cyclic.
	(Interestingly, if $Q_\text{hk}$ is defined as in Refs.~\cite{Komatsu2008,Komatsu2010}, then for cyclic paths this integral is described in terms of a geometric phase~\cite{Sagawa2011}.)
	
	Recall that reversible equilibrium processes, which satisfy $\Delta S + \int\mathrm{d}t \, \beta \langle \dot{Q}\rangle = 0$, are characterized by zero entropy production in the universe: any change in the system's entropy is balanced by a compensating change in its surroundings.
	By analogy, in quasistatic nonequilibrium processes, which satisfy Eq.~\ref{eq:mClausius_rev},
	the entropy change of the system, $\Delta S$, is balanced by the {\it excess} entropy produced in the reservoir, $\int \mathrm{d}t \, \beta \langle \dot{Q}_\text{ex} \rangle$. 
	(The total entropy production in the reservoir diverges in the quasistatic limit, $\int \mathrm{d}t \, \beta \langle \dot{Q} \rangle\rightarrow\infty$, due to the continual flow of housekeeping heat.)
	Moreover, just as a system remains arbitrarily close to equilibrium during a reversible processes, a system undergoing a quasistatic nonequilibrium transition remains arbitrarily close to the nonequilibrium steady state ($\Delta{\bf p} \propto \epsilon$).
	In both cases, equilibrium and nonequilibrium, the system retraces its path in the reverse order when it is subjected to the reverse process ${\boldsymbol \lambda}: {\bf A} \leftarrow {\bf B}$; in this sense, there is no hysteresis.
	In view of these parallels, it is natural to think of quasistatic nonequilibrium processes as the nonequilibrium analogues of reversible equilibrium processes, as suggested by Oono and Paniconi~\cite{Oono1998}.

	Let us now move beyond the quasistatic limit, by including the $n = 2$ term of Eq.~\ref{eq:Series} in the analysis. 
	Starting with Eq.~\ref{eq:mExcess}, we obtain
	\begin{equation}
	\label{eq:Slow}
	\beta \langle \dot{Q}_\text{ex} \rangle = - \frac{dS}{dt}  + \sum_{i, j} R_{ij}  \ln{\pi_i} \sum_{k, l} R^+_{jk} \frac{d}{dt} \left( R^+_{kl} \dot{\pi_l} \right)
	\end{equation}
	in place of Eq.~\ref{eq:reversible}.
	Integrating with respect to time, we obtain, after some simplifying steps (see SI),
	\begin{align}
	\label{eq:Slow1}
	\Delta S \, + \, \int \mathrm{d}t \,  \beta \langle \dot{Q}_\text{ex} \rangle & \nonumber \\ 
	= \Delta \sum_{i,j} \ln{\pi_i} R^+_{ij} \dot{\pi_j} & - \int \mathrm{d}t \sum_{i, j} \pi_j \frac{d \ln{\pi_i}}{dt} R^+_{ij} \frac{d \ln{\pi_j}}{dt}.
	\end{align}
	If we now assume that $d{\boldsymbol \lambda}/dt=0$ at the start and end of the process, then the first term on the right of Eq.~\ref{eq:Slow1} vanishes. 
	As the steady states ${\boldsymbol \pi}$ are determined by the parameters ${\boldsymbol \lambda}$, we can rewrite Eq.~\ref{eq:Slow1} in the form
	\begin{equation}
	\label{eq:Slow2}
	\begin{split}
	\Delta S + \int \mathrm{d}t \, \beta \langle \dot{Q}_\text{ex} \rangle &= \int {\mathrm d}t\, \dot{{\boldsymbol \lambda}}^T {\xi}({\boldsymbol \lambda}) \dot{{\boldsymbol \lambda}} \\
	&= \int {\mathrm d}t\, \dot{{\boldsymbol \lambda}}^T {\zeta}({\boldsymbol \lambda}) \dot{{\boldsymbol \lambda}} ,
	\end{split}
	\end{equation}
where $\zeta = (\xi + \xi^T)/2$ is the symmetric part
	 of a matrix $\xi({\boldsymbol\lambda})$ whose elements are
	\begin{equation}
	\label{eq:Metric}
	{\xi}_{\mu \nu}  = -  \sum_{i, j} \pi_j\frac{\partial \ln{\pi_i }}{\partial \lambda_\nu}R^+_{ij} \frac{\partial \ln{\pi_j }}{\partial \lambda_\mu}.
	\end{equation}
	Equation~\ref{eq:Slow2} provides the leading correction to Eq.~\ref{eq:mClausius_rev}, and is the counterpart of analogous results for slow transitions between equilibrium states~\cite{Speck2004,Sivak2012,Hoppenau2013}.

	We now derive a Green-Kubo relation for the elements of the matrix $\zeta({\boldsymbol \lambda})$.  
	Let us define a set of observables
	\begin{equation}
	\label{eq:StateFunc}
	F_i^\mu({\boldsymbol\lambda}) = \frac{\partial \ln{\pi_i}({\boldsymbol \lambda})}{\partial \lambda_\mu}
	\quad,\quad \mu = 1,\cdots K.
	\end{equation}
	When the system is in the steady state ${\boldsymbol \pi}^{\boldsymbol \lambda}$, its microstate $i(t) \in \{1, 2, \ldots N\}$ fluctuates in time, hence so does each $F^\mu(t) \equiv F^\mu_{i(t)}$, around a mean value $\langle F^\mu\rangle_{\boldsymbol \lambda} = 0$. 
	Letting $\langle F^\mu(0) F^\nu(t) \rangle_{{\boldsymbol \lambda}}$ denote a correlation function evaluated in the nonequilibrium steady state, the matrix elements $\zeta_{\mu\nu}$ can be rewritten as (see SI for details):
	\begin{equation}
	\label{eq:FDT}
	\zeta_{\mu \nu}({\boldsymbol \lambda}) = \frac{1}{2} \int_{-\infty}^{+\infty} \mathrm{d}t \, \langle F^\mu(0) F^\nu(t) \rangle_{{\boldsymbol \lambda}} .
	\end{equation}
	This result relates an excess dissipation coefficient $\zeta_{\mu\nu}$ to stationary fluctuations in the nonequilibrium steady state.
	(Analogously, for near-equilibrium transitions the friction tensor is determined by equilibrium fluctuations~\cite{Sivak2012}.)
	We emphasize that the steady state in Eq.~\ref{eq:FDT} may be far from thermal equilibrium.
	
	As shown by Prost {\it et al}~\cite{Prost2009}, and for general Markov processes by H\" anggi and Thomas~\cite{Hanggi1982}, an expression similar to Eq.~\ref{eq:FDT} describes the linear response of a system to small perturbations around a given steady state.
	By contrast, our analysis applies to slow transitions between two steady states that may differ substantially.
	
	The left side of Eq.~\ref{eq:mClausius} (or Eq.~\ref{eq:Slow2}) is the ensemble average of a quantity identified by Esposito {\it et al}~\cite{Esposito2007,Esposito2010} as the {\it nonadiabatic} component of entropy production:
	\begin{equation}
	\label{eq:Deltasna}
	\Delta s_{\rm na} = -\ln\pi_{i(\tau)}^{\bf B} + \ln\pi_{i(0)}^{\bf A} + \int \mathrm{d}t \, \beta \dot{Q}_\text{ex} ,
	\end{equation}
	where $i(t)$ is the microstate of the system during a single realization of the process.
	For slow driving the mean and variance of $\Delta s_{\rm na}$ satisfy (see SI)
	\begin{equation}
	\label{eq:Gaussian}
	\langle \Delta s_{\rm na} \rangle = \frac{1}{2} \sigma_{\Delta s_{\rm na}}^2 .
	\end{equation}
	Let us place this result in context.
	If the dynamics satisfy detailed balance and temperature is constant, Eq.~\ref{eq:Deltasna} reduces to $\Delta s_{\rm na} = \beta(W - \Delta F)$, where $W$ is the work performed on the system and $\Delta F$ is the free energy difference between equilibrium states ${\bf A}$ and ${\bf B}$.
	Eq.~\ref{eq:Gaussian} then becomes
	\begin{equation}
	\label{eq:GaussianEq}
	\langle W \rangle = \Delta F + \frac{\beta}{2} \sigma_W^2.
	\end{equation}
	This fluctuation-dissipation relation for isothermal, near-equilibrium processes was originally proposed by Hermans~\cite{Hermans1991} and Wood~\cite{Wood1991}, using a Gaussian assumption for the work distribution, and more recently has been obtained using projection operator techniques by Speck and Seifert~\cite{Speck2004} and multiple time-scale analysis by Hoppenau and Engel~\cite{Hoppenau2013}.
	Our result, Eq.~\ref{eq:Gaussian}, generalizes Eq.~\ref{eq:GaussianEq} to slow transitions between nonequilibrium steady states.
	Using the techniques of Refs.~\cite{Speck2004, Hoppenau2013}, one can show that for such transitions, typical values of $\Delta s_{\rm na}$ follow a Gaussian distribution. 

	Since the matrix $\zeta({\boldsymbol \lambda})$ is positive semidefinite (see SI), it provides a metric in ${\boldsymbol \lambda}$-space, related to nonadiabatic entropy production for slow processes. 
	Specifically, for a protocol ${\boldsymbol \lambda}(t)$ describing a contour ${\cal C}$ in parameter space, we define a length $l_{\cal C}$ by the contour integral
	\begin{equation}
	\label{eq:Length}
l_{\cal C} = \int_{\cal C}  \sqrt{{\mathrm d}{\boldsymbol \lambda}^T {\zeta}({\boldsymbol \lambda}) \,  {\mathrm d}{\boldsymbol \lambda}}.
	\end{equation}
	By the Cauchy-Schwarz inequality we have
	\begin{equation}
	\label{eq:CSIneq}
	\int {\mathrm d}t\, \dot{{\boldsymbol \lambda}}^T {\zeta}({\boldsymbol \lambda}) \dot{{\boldsymbol \lambda}} \geq \frac{l_{\cal C}^2}{\Delta t},
	\end{equation}
where $\Delta t$ is the duration of the transition.
	If the driving is sufficiently slow that Eq.~\ref{eq:Slow2} applies, then the right side of Eq.~\ref{eq:CSIneq} provides a lower bound for the nonadiabatic entropy production: $\langle\Delta s_{\rm na}\rangle \geq l_{\cal C}^2/\Delta t$.
	This result generalizes a similar structure for transitions between equilibrium states, which has been used to determine optimal (minimally dissipative) protocols ${\boldsymbol \lambda}(t)^\text{opt}$ for fixed end-points ${\bf A}$ and ${\bf B}$ and duration $\Delta t$~\cite{Salamon1983, Crooks2007, Sivak2012, Zulkowski2012}.
	
	Finally, following Sivak and Crooks~\cite{Sivak2012}, we note that the metric $\zeta({\boldsymbol \lambda})$ is the Hadamard (term by term) product 
	\begin{equation}
	\label{eq:Hadamard_NESS}
	{\zeta}({\boldsymbol \lambda}) = {{\cal I}} ({\boldsymbol \lambda}) \circ {\tau}({\boldsymbol \lambda}),
	\end{equation}	
of the Fisher information matrix~\cite{Cover1991} of the steady state distribution $\boldsymbol \pi^{\boldsymbol \lambda}$,  
	\begin{equation}
	\label{eq:Fisher_NESS}
	{\cal I}_{\mu \nu} ({\boldsymbol \lambda}) = \sum_i \pi_i \left( \partial_\mu \ln{\pi_i} \right) \left( \partial_\nu \ln{\pi_i} \right) = \langle F^\mu F^\nu \rangle_{\boldsymbol \lambda},
	\end{equation}
and a matrix generalization $\tau({\boldsymbol \lambda})$ of the relaxation time,
	\begin{equation}
	\label{eq:Relaxation_NESS}
	\tau_{\mu \nu} = \int_0^\infty {\mathrm d}t \, \frac{\langle F^\mu(t) F^\nu(0) \rangle_{\boldsymbol \lambda}}{\langle F^\mu F^\nu \rangle_{\boldsymbol \lambda}}.
	\end{equation}
	In the framework of finite-time thermodynamics~\cite{Andresen1984, Nulton1985,Andresen2011}, built on the metric geometry of equilibrium thermodynamics~\cite{Weinhold1975, Ruppeiner1979, Salamon1980, Gilmore1984, Andresen1988, Feng2008}, the Fisher-information matrix ${\cal I}({\boldsymbol \lambda})$ alone is assumed to dictate the average dissipation of heat during a thermodynamic process. 
	The current work, in agreement with Refs.~\cite{Sivak2012, Zulkowski2012, Zulkowski2013}, emphasizes the importance of relaxation dynamics, via the matrix $\tau({\boldsymbol \lambda})$. 

	This material is based upon work supported by the U.S. Army Research Laboratory and the U.S. Army Research Office under contract number W911NF-13-1-0390.

	\newpage\newpage
      \renewcommand{\thepage}{S\arabic{page}}
      \setcounter{page}{1}
      \renewcommand{\theequation}{S\arabic{equation}}
      \setcounter{equation}{0}

	
	\section{Supplementary information}
	\vskip 0.1in

	We begin by deriving Eq.~\ref{eq:SMPInverse3} below, which we will later use to derive Eq.~23 of the main text.
	We note that Eq.~11, which uniquely specifies the matrix ${\cal R}^+$,\footnote{
	In fact, there is some redundancy: given Eq.~11a, the two equations appearing in Eq.~11b are equivalent to one another.} can be written as a set of four equations:
	\begin{eqnarray}
	\label{eq:Rplus1}
	{\cal R} {\cal R}^+ & = & {\bf I} - {\boldsymbol\pi} {\bf 1}^T \\
	\label{eq:Rplus2}
	{\cal R}^+ {\cal R} & = & {\bf I} - {\boldsymbol\pi} {\bf 1}^T \\
	\label{eq:Rplus3}
	{\cal R}^+ {\boldsymbol \pi} & = & {\bf 0} \\
	\label{eq:Rplus4}
	{\bf 1}^T {\cal R}^+ & = & {\bf 0}^T
	\end{eqnarray}
where ${\bf I}$ is the identity matrix. 
	In the following, we show that the integral
	\begin{equation}
	{\cal M} \equiv \int_0^\infty \mathrm{d} t \, e^{{\cal R} t} ({\boldsymbol \pi} {\bf 1}^T - {\bf I})
	\end{equation}
satisfies all these defining relations of ${\cal R}^+$, thereby constituting an exact expression of the latter. 
	In these calculations, ${\cal R}$ and ${\boldsymbol\pi}$ are fixed (i.e.\ time-independent).

	{\it Eq.~\ref{eq:Rplus1}}.
	Multiplying ${\cal M}$ by ${\cal R}$ on the left, we get	
	\begin{equation}
	\label{eq:SMPInverse1}
	\begin{split}
	{\cal R} {\cal M} & =  {\cal R} \int_0^\infty \mathrm{d} t \, e^{{\cal R} t} ({\boldsymbol \pi} {\bf 1}^T - {\bf I}) \\
	& = \int_{t=0}^{t=\infty} \mathrm{d} \left( e^{{\cal R} t} \right) ({\boldsymbol \pi} {\bf 1}^T - {\bf I}) \\
	& =  ({\boldsymbol \pi} {\bf 1}^\text{T} - {\bf I})^2 \\
	& =  {\bf I} - {\boldsymbol \pi} {\bf 1}^T.
	\end{split}
	\end{equation}
	The third line follows from the relaxation property: $\lim_{t \rightarrow \infty} \exp{({\cal R} t)} {\bf p}(0) = {\boldsymbol \pi}$ for any normalized ${\bf p}(0)$,
	hence $\lim_{t \rightarrow \infty} \exp{({\cal R} t)} = {\boldsymbol \pi} {\bf 1}^T$.
	The last line follows from the normalization ${\bf 1}^T {\boldsymbol \pi} = 1$ and a cancellation of terms in the product.

	{\it Eq.~\ref{eq:Rplus2}}.	
	Multiplying ${\cal M}$ by ${\cal R}$ on the right, we get	
	\begin{equation}
	\label{eq:SMPInverse2}
	\begin{split}
	{\cal M} {\cal R} & =  \int_0^\infty \mathrm{d} t \, e^{{\cal R} t} ({\boldsymbol \pi} {\bf 1}^T - {\bf I}) {\cal R} \\
	& =  \int_0^\infty \mathrm{d} t \, e^{{\cal R} t} {\cal R} ({\boldsymbol \pi} {\bf 1}^T - {\bf I}) \\
	& =  \int_{t=0}^{t=\infty} \mathrm{d} \left( e^{{\cal R} t} \right) ({\boldsymbol \pi} {\bf 1}^T - {\bf I}) \\
	& =  {\bf I} - {\boldsymbol \pi} {\bf 1}^T.
	\end{split}
	\end{equation}
	The second line utilizes the relations: ${\bf 1}^T {\cal R} = {\bf 0}^T$ and ${\cal R} {\boldsymbol \pi} = {\bf 0}$. 
	The last line follows from Eq.~\ref{eq:SMPInverse1}.

	{\it Eq.~\ref{eq:Rplus3}}. Multiplying ${\cal M}$ by ${\boldsymbol \pi}$ we get
	\begin{equation}
	\begin{split}
	{\cal M} {\boldsymbol \pi} & = \int_0^\infty \mathrm{d} t \, e^{{\cal R} t} ({\boldsymbol \pi} {\bf 1}^T - {\bf I}) {\boldsymbol \pi} \\
	& = \int_0^\infty \mathrm{d} t \, e^{{\cal R} t}  ({\boldsymbol \pi} - {\boldsymbol \pi}) = {\bf 0}.
	\end{split}
	\end{equation}
	The second line follows from the normalization condition ${\bf 1}^T {\boldsymbol \pi} = 1$.

	{\it Eq.~\ref{eq:Rplus4}}. Multiplying ${\cal M}$ by ${\bf 1}^T$ we get
	\begin{equation}
	\label{eq:SMPInverse4}
	\begin{split}
	{\bf 1}^T {\cal M} & = \int_0^\infty \mathrm{d} t \, {\bf 1}^T e^{{\cal R} t} ({\boldsymbol \pi} {\bf 1}^T - {\bf I}) \\
	& = \int_0^\infty \mathrm{d} t \, {\bf 1}^T ({\boldsymbol \pi} {\bf 1}^T - {\bf I}) \\
	& = \int_0^\infty \mathrm{d} t \, ({\bf 1}^T - {\bf 1}^T)  = {\bf 0}^T.
	\end{split}
	\end{equation}
	The second line follows from the stationary relation ${\bf 1}^T e^{{\cal R}t} = {\bf 1}^T$. 
	The third line utilizes normalization condition of ${\boldsymbol \pi}$. 
	Comparing Eqs.~\ref{eq:SMPInverse1}--\ref{eq:SMPInverse4} with Eqs.~\ref{eq:Rplus1}--\ref{eq:Rplus4}, we see that ${\cal M}$ satifies all the defining relations of ${\cal R}^+$, leading to the identification
	\begin{equation}
	\label{eq:SMPInverse3}
	{\cal R}^+ = {\cal M} = \int_0^\infty \mathrm{d} t \, e^{{\cal R} t} ({\boldsymbol \pi} {\bf 1}^T - {\bf I}). 
	\end{equation}

	\subsection{Derivation of Eq.~19}

	We start with Eq.~18:
	\begin{equation}
	\label{eq:M16}
	\beta \langle \dot{Q}_\text{ex} \rangle  =  - \frac{dS}{dt} + \sum_{i j k l} R_{ij} \ln{\pi_i} R^+_{jk} \frac{d}{dt} \left(R^+_{kl} \dot{\pi}_l \right).
	\end{equation}
	Summing over the index $j$ and using Eq.~11a, the second term on the right becomes
	\begin{equation}
	\label{eq:SExcessHeat}
	\sum_{i k l} \! \! \left( \delta_{ik} - \pi_i \right) \ln{\pi_i} \frac{d}{dt}\left( R^+_{kl} \dot{\pi} _l \right).
	\end{equation}
	Summing over $i$ and using $S = - \sum_i \pi_i \ln{\pi_i}$, Eq.~\ref{eq:SExcessHeat} becomes
	\begin{equation}
	\label{eq:SExcessHeat3}
	\sum_{kl} ( \ln\pi_k + S ) \frac{d}{dt}\left( R^+_{kl} \dot{\pi} _l \right) .
	\end{equation}
	The second term in Eq.~\ref{eq:SExcessHeat3} (containing $S$) vanishes upon summing over $k$, since ${\bf 1}^T{\cal R}^+ = {\bf 0}^T$ (Eq.~11b).
	Combining results, we get
	\begin{equation}
	\label{eq:SExcessHeat4}
	\beta \langle \dot{Q}_\text{ex} \rangle = - \frac{dS}{dt}  + \sum_{k l} \ln{\pi_k} \frac{d}{dt}\left( R^+_{kl} \dot{\pi} _l \right).
	\end{equation}
	Integrating both sides with respect to time leads us to Eq.~19, after integration by parts.

	\subsection{Derivation of Eq.~23}
	
	From the definition of $F_i^\mu$ (Eq.~22), we have
	\begin{equation}
	\label{eq:SFDT1}
	\int_0^\infty \! \! \! \!  \mathrm{d}t \, \langle F^\mu(0) F^\nu(t) \rangle_{\boldsymbol \lambda} = 
	\int_0^\infty \! \! \! \! \mathrm{d}t \, \sum_{i j}  \! \pi_j  \frac{\partial \ln{\pi_i}}{\partial \lambda_\nu} \left( e^{{\cal R} t} \right)_{ij} 
	\frac{\partial \ln{\pi_j}}{\partial \lambda_\mu} ,
	\end{equation}
	with all quantities evaluated at fixed ${\boldsymbol\lambda}$.
	We can rewrite the right hand side as
	\begin{equation}
	\begin{split}
	\label{eq:temp}
	& \int_0^\infty \! \! \! \! \mathrm{d}t \, \sum_{i j}  \! \pi_j  \frac{\partial \ln{\pi_i}}{\partial \lambda_\nu} \left( e^{{\cal R} t} \right)_{ij} 
	\frac{\partial \ln{\pi_j}}{\partial \lambda_\mu} \\
	= & \int_0^\infty \! \! \! \! \mathrm{d}t \, \sum_{i}   \frac{\partial \ln{\pi_i}}{\partial \lambda_\nu} \left( e^{{\cal R} t} \frac{\partial {\boldsymbol \pi}}{\partial \lambda_\mu} \right)_i \\
	= & \int_0^\infty \! \! \! \! \mathrm{d}t \, \sum_{i}   \frac{\partial \ln{\pi_i}}{\partial \lambda_\nu} \left[ \left( e^{{\cal R} t} - {\boldsymbol \pi} {\bf 1}^T \right) \frac{\partial {\boldsymbol \pi}}{\partial \lambda_\mu} \right]_i \\
	= & \int_0^\infty \! \! \! \! \mathrm{d}t \, \sum_{i} \frac{\partial \ln{\pi_i}}{\partial \lambda_\nu} \left[ e^{{\cal R} t} \left( {\bf I} - {\boldsymbol \pi} {\bf 1}^T \right) \frac{\partial  {\boldsymbol \pi}}{\partial \lambda_\mu} \right]_i \\
	= & \int_0^\infty \! \! \! \! \mathrm{d}t \, \sum_{i j} \pi_j  \frac{\partial \ln{\pi_i}}{\partial \lambda_\nu} \left[ e^{{\cal R} t} \left( {\bf I} - {\boldsymbol \pi} {\bf 1}^T \right) \right]_{ij} \frac{\partial  \ln{\pi_j}}{\partial \lambda_\mu} \\
	= - & \sum_{i j} \pi_j  \frac{\partial \ln{\pi_i}}{\partial \lambda_\nu} {\cal R}^+_{ij} \frac{\partial  \ln{\pi_j}}{\partial \lambda_\mu},
	\end{split}
	\end{equation}
where in the third line we have used ${\bf 1}^T \cdot (\partial {\boldsymbol \pi} / \partial \lambda_\mu ) = 0$, which follows from the normalization condition ${\bf 1}^T {\boldsymbol \pi} = 1$; in the fourth line we have used the stationarity condition $e^{{\cal R} t} {\boldsymbol \pi} = {\boldsymbol \pi}$; and in the last line we have used Eq.~\ref{eq:SMPInverse3}. Combining Eqs.~\ref{eq:SFDT1} and \ref{eq:temp}, we get
	\begin{equation}
	\label{eq:SFDT2}
	\begin{split}
	\int_0^\infty \! \! \! \!  \mathrm{d}t \, \langle F^\mu(0) F^\nu(t) \rangle_{\boldsymbol \lambda} & = - \sum_{i j} \pi_j \frac{\partial \ln{\pi_i}}{\partial \lambda_\nu} R^+_{ij} \frac{\partial \ln{\pi_j}}{\partial \lambda_\mu} \\
	&= \xi_{\mu\nu} \quad ,
	\end{split}
	\end{equation}
	using Eq.~21.
	We also have
	\begin{equation}
	\label{eq:SFDT3}
	\int_{-\infty}^0 \! \! \! \!  \mathrm{d}t \, \langle F^\mu(0) F^\nu(t) \rangle_{\boldsymbol \lambda}
	= \int_0^\infty \! \! \! \!  \mathrm{d}t \, \langle F^\mu(t) F^\nu(0) \rangle_{\boldsymbol \lambda}
	= \xi_{\nu\mu}
	\end{equation}
	using Eq.~\ref{eq:SFDT2}.
	Combining Eqs.~\ref{eq:SFDT2} and \ref{eq:SFDT3} gives us the desired result:
	\begin{equation}
	\label{eq:GreenKubo}
	\zeta_{\mu\nu} \equiv \frac{1}{2} \left( \xi_{\mu\nu} + \xi_{\nu\mu} \right) = \frac{1}{2} \int_{-\infty}^{+\infty} \! \! \! \!  \mathrm{d}t \, \langle F^\mu(0) F^\nu(t) \rangle_{\boldsymbol \lambda}.
	\end{equation}

	\subsection{Derivation of Eq.~25}

	By Eq.~8, the excess heat associated with a transition from state $j$ to state $i$ is equal to $\beta^{-1}\ln(\pi_i/\pi_j)$.
	Thus the total excess heat dissipated during a single realization of the process is given by
	\begin{equation}
	\label{eq:SExcessHeat5}
	Q_\text{ex} = \beta^{-1} \sum_{t_n} \ln{\left[\frac{\pi_{i(t_n+)}}{\pi_{i(t_n -)}} \right]},
	\end{equation}
	where $i(t_n-)$ and $i(t_n+)$ denote the states of the system just before and just after a transition at time $t_n$.
	Eq.~24 now becomes, with $\tau$ as the duration of the process,
	\begin{eqnarray}
	\label{eq:SNonAd}
	\Delta s_{\rm na} &=& - \ln \pi_{i(\tau)}^{\bf B} + \ln \pi_{i(0)}^{\bf A} + \sum_{t_n} \ln{\left[\frac{\pi_{i(t_n+)}}{\pi_{i(t_n -)}} \right]} \nonumber\\
	&=& - \int_0^\tau  \mathrm{d}t \,   \sum_\mu \dot{\lambda}_\mu(t) \left[ \frac{\partial \ln{\pi_{i(t)}^{\boldsymbol \lambda}}}{\partial \lambda_\mu} \right]_{{\boldsymbol \lambda}(t)} \nonumber\\
	& = & - \int_0^\tau  \mathrm{d}t \,   \sum_\mu \dot{\lambda}_\mu(t) \bar F^\mu(t),
	\end{eqnarray}
	where $\bar F^\mu(t) \equiv F_{i(t)}^\mu({\boldsymbol\lambda}(t))$.
	In going from the first line to the second, we have used the fact that the net change in the value of $\ln\pi_{i(t)}^{{\boldsymbol\lambda}(t)}$, from $t=0$ to $t=\tau$, is a sum of contributions  due to: (1) discrete changes in $i(t)$ at the transitions, and (2) the continuous variation of ${\boldsymbol\lambda}(t)$ between transitions.
	
	Over an ensemble of realizations, we have
	\begin{equation}
	\label{eq:meanSquare_1}
	\langle (\Delta s_{\rm na})^2 \rangle = \int_0^\tau \!\!\!\! {\rm d}t \, \int_0^\tau \!\!\!\! \mathrm{d}t^\prime \, \sum_{\mu \nu} \dot{\lambda}_\mu(t) \dot{\lambda}_\nu(t^\prime) \langle \bar F^\mu(t) \bar F^\nu(t^\prime) \rangle.
	\end{equation}
	Let us now evaluate this expression in the limit of slow driving, where there is a separation of time scales between the slow variation of ${\boldsymbol\lambda}(t)$ and the fast evolution of $\bar{\bf F}(t)$ (due to rapid transitions between states).
	As in the main text, it is convenient to think in terms of a small parameter $\epsilon\propto\vert\dot{\boldsymbol\lambda}\vert$, so that changes in ${\boldsymbol\lambda}(t)$ occur on time scales of order $\epsilon^{-1}$, while changes in $\bar{\bf F}(t)$ occur on time scales of order unity.
	The correlation function $\langle \bar F^\mu(t) \bar F^\nu(t^\prime) \rangle$ decays with $\vert t^\prime-t\vert$, on a time scale of order unity.
	As a result, to leading order we can replace this correlation function with one that is evaluated in a given nonequilibrium steady state:
	\begin{equation}
	\label{eq:approxCF}
	\langle \bar F^\mu(t) \bar F^\nu(t^\prime) \rangle \approx \langle F^\mu(t) F^\nu(t^\prime) \rangle_{\boldsymbol\lambda} \equiv C^{\mu\nu}(s;{\boldsymbol\lambda}),
	\end{equation}
	with $s = t^\prime - t$ and ${\boldsymbol\lambda} = {\boldsymbol\lambda}(t)$ in Eq.~\ref{eq:approxCF}.
	Here as in the main text, $F^\mu(t)$ is evaluated along a trajectory generated at fixed ${\boldsymbol\lambda}$, unlike $\bar F^\mu(t)$ -- defined above -- which is evaluated along a trajectory evolving under the slow variation of the external parameters.
	Eq.~\ref{eq:meanSquare_1} now becomes
	\begin{equation}
	\label{eq:meanSquare_2}
	\langle (\Delta s_{\rm na})^2 \rangle \approx \sum_{\mu \nu} \int_0^\tau \!\!\!\! {\rm d}t \, \dot{\lambda}_\mu(t) \dot{\lambda}_\nu(t) \int_{-t}^{\tau-t} \!\!\!\! \mathrm{d}s \, C^{\mu\nu}(s;{\boldsymbol\lambda}(t)). \nonumber
	\end{equation}
	Since $\tau\propto\epsilon^{-1}$, we conclude that for most values of $t$ between $0$ and $\tau$, both $t$ and $\tau-t$ are much larger than the time scale over which the correlation function decays.
	Hence to leading order in $\epsilon$ we can write
	\begin{eqnarray}
	\label{eq:meanSquare_3}
	\langle (\Delta s_{\rm na})^2 \rangle &\approx& \sum_{\mu \nu} \int_0^\tau \!\!\!\! {\rm d}t \, \dot{\lambda}_\mu(t) \dot{\lambda}_\nu(t) \int_{-\infty}^{+\infty} \!\!\!\! \mathrm{d}s \, C^{\mu\nu}(s;{\boldsymbol\lambda}(t)) \nonumber \\
	&=& 2 \int_0^\tau \!\!\!\! {\rm d}t \, \dot{\boldsymbol\lambda}^T \zeta({\boldsymbol\lambda}) \dot{\boldsymbol\lambda}
	\end{eqnarray}
	using Eq.~\ref{eq:GreenKubo}.
	Now note that Eq.~20 can be written as
	\begin{equation}
	\langle\Delta s_{\rm na}\rangle = \int_0^\tau \!\!\!\! {\rm d}t \, \dot{\boldsymbol\lambda}^T \zeta({\boldsymbol\lambda}) \dot{\boldsymbol\lambda} \quad .
	\end{equation}
	Since $\dot{\boldsymbol\lambda}\propto\epsilon$ and $\tau\propto\epsilon^{-1}$, both $\langle (\Delta s_{\rm na})^2\rangle$ and $\langle\Delta s_{\rm na}\rangle$ scale as $\epsilon$, which implies that
	\begin{equation}
	\sigma_{\Delta s_{\rm na}}^2 \equiv \langle (\Delta s_{\rm na})^2 \rangle - \langle\Delta s_{\rm na}\rangle^2 \approx \langle (\Delta s_{\rm na})^2 \rangle.
	\end{equation}
	Hence to leading order we have $\langle\Delta s_{\rm na}\rangle = (1/2) \sigma_{\Delta s_{\rm na}}^2$.

	\vskip 0.7in

	\subsection{The matrix $\zeta({\boldsymbol\lambda})$ is positive semidefinite}
	
	To establish this result, let $i(t)$ denote a trajectory evolving under the stationary dynamics at fixed $\boldsymbol\lambda$, let $\Delta t > 0$ be an interval of time, let $\{ a_1, a_2, \cdots a_K \}$ denote a set of real values,	and consider a quantity
	\begin{equation}
	Y = \int_0^{\Delta t} {\rm d}t \, \sum_\mu a_\mu F_{i(t)}^\mu .
	\end{equation}
	Squaring the value of $Y$ and averaging over an ensemble of steady-state trajectories, we get
	\begin{eqnarray}
	\langle Y^2 \rangle &=& \sum_{\mu\nu} a_\mu a_\nu
	\int_0^{\Delta t} \!\!\!\! {\rm d}t \, \int_0^{\Delta t} \!\!\!\! \mathrm{d}t^\prime \, \langle F^\mu(t) F^\nu(t^\prime) \rangle_{{\boldsymbol\lambda}} \nonumber \\
	&=& \sum_{\mu\nu} a_\mu a_\nu \int_0^{\Delta t} \!\!\!\! {\rm d}t \, \int_{-t}^{\Delta t-t} \!\!\!\! \mathrm{d}s \, C^{\mu\nu}(s;{\boldsymbol\lambda}) \nonumber \\
	&=& \sum_{\mu\nu} a_\mu a_\nu \int_{-\Delta t}^{+\Delta t} \!\!\!\! {\rm d}s \, \left( \Delta t - \vert s\vert \right) C^{\mu\nu}(s;{\boldsymbol\lambda}) 
	\end{eqnarray}
	If we now divide both sides by $\Delta t$ and consider the limit $\Delta t\rightarrow\infty$, we arrive at
	\begin{eqnarray}
	\lim_{\Delta t\rightarrow\infty} \frac{\langle Y^2\rangle}{\Delta t} &=& \sum_{\mu\nu} a_\mu a_\nu \int_{-\infty}^{+\infty} \!\!\!\! {\rm d}s \, C^{\mu\nu}(s;{\boldsymbol\lambda}) \nonumber \\
	&=& 2 \sum_{\mu\nu} a_\mu \zeta_{\mu\nu} a_\nu
	\end{eqnarray}
	Since the left side is non-negative for any choice of $\{ a_1, a_2, \cdots a_K \}$, we conclude that $\zeta({\boldsymbol\lambda})$ must be positive semidefinite.

	\end{document}